\documentstyle[11pt]{article}
\begin{document}
\title{Pair creation rate for $U(1)^2$ black holes}
\author{Simon F. Ross \\
 {\it Department of Applied Mathematics and Theoretical Physics} \\
 {\it University of Cambridge, Silver St., Cambridge CB3 9EW} \\
 S.F.Ross@damtp.cam.ac.uk}
\date{\today \\ DAMTP/R-95/47}
\maketitle

\begin{abstract}
  I consider a truncation of low-energy string theory which contains
  two $U(1)$ gauge fields. After making some general comments on the
  theory, I describe a previously-obtained instanton for the pair
  creation of black holes when both gauge fields are non-zero, and
  obtain the pair creation rate by calculating its action. This
  calculation agrees qualitatively with the earlier calculation of the
  pair creation rate for black holes in Einstein-Maxwell theory. That
  is, the pair creation is strongly suppressed in realizable
  circumstances, and it reduces to the Schwinger result in the
  point-particle limit. The pair creation of non-extreme black holes
  is enhanced over that of extreme black holes by $e^{{\cal
      A}_{bh}/4}$.
\end{abstract}

\pagebreak

\section{Introduction}
\label{intro}

The study of black hole pair creation is of considerable interest for
the exploration of quantum gravity. Like black hole evaporation, it
represents a truly quantum gravitational process, being classically
completely forbidden. At the same time, it is easy to achieve a
physical understanding of what is happening; there is a strong analogy
with the creation of particle-antiparticle pairs in quantum field
theory, as can be seen from the fact that the black hole pair creation
rate reduces to the particle-antiparticle rate in the limit of small
black holes. If we trust this analogy to particle physics, the pair
creation rate should depend on the number of accessible states for the
black hole, so we can find out how many states the black holes should
have by studying pair creation (although we can't find out what those
states are).  Black hole pair creation involves topology change, and
this suggests that including the effects of topology change will be
important to a proper understanding of quantum gravity.

Because of the topology change, black hole pair creation is studied in
the path-integral approach to quantum gravity, by finding a suitable
instanton (that is, a solution of the classical equations of motion
with Euclidean signature) which describes the transition from the
given initial to final data. Most of the work to date has focussed on
the pair creation of charged black holes in a background
electromagnetic field, both in Einstein-Maxwell theory \cite{ggs} and
in a generalisation of this theory which includes a dilaton
\cite{dgkt,dggh}, whose action is
\begin{equation} \label{Faction}
I = - \frac{1}{16\pi} \int_M ( R - 2 \partial^{\mu}\phi
\partial_{\mu} \phi -e^{-2a\phi} F^2 ) -\frac{1}{
8\pi}\oint_{\partial M} (K-K_0).
\end{equation}

The present paper is concerned with a different generalisation of
Einstein-Maxwell theory, to include two $U(1)$ gauge fields and a
dilaton, called the $U(1)^2$ theory. This is a somewhat more
typical example of a low-energy effective theory arising from
superstring theory, as the compactification of the extra dimensions
will typically give an effective theory with a large number of $U(1)$
gauge fields. As I will argue in Sec. \ref{dsec}, the most
appropriate effective action for this theory is  \cite{Kalloshcens}:
\begin{equation} \label{Kaction}
I_{SO(4)} = -\frac{1}{16 \pi} \int_M ( R - 2 \partial^{\mu}\phi
\partial_{\mu} \phi  -(e^{2\phi} \tilde{F}^2 +
e^{-2\phi} G^2)) -\frac{1}{
8\pi}\oint_{\partial M} (K-K_0),
\end{equation}
where $\tilde{F}_{\mu\nu}$ and $G_{\mu\nu}$ are the two $U(1)$ gauge
fields. This theory is a consistent truncation of low-energy heterotic
string theory \cite{Kalloshcens}. One of the advantages of this
truncation is that it includes the Einstein-Maxwell theory as a
special case, when $\phi=0$ and $\tilde{F}_{\mu\nu} = G_{\mu\nu}$.
That is, Einstein-Maxwell is also a consistent truncation of string
theory. The truncation (\ref{Kaction}) can also be derived from the
$SO(4)$ version of $N=4$ supergravity \cite{cremmer}.  It also
includes the action (\ref{Faction}) with $a=1$, when one of the gauge
fields vanishes.

There are two duality symmetries in the $U(1)^2$ theory; one of them
is a generalisation of the usual electric-magnetic duality, while the
other is trivial on the Einstein-Maxwell solutions.  Charged black
hole solutions of (\ref{Kaction}) were found by Gibbons \cite{Gary}.
These solutions include the Reissner-Nordstr\"om metrics when the two
gauge charges are equal, so the Reissner-Nordstr\"om solutions
correspond to dyonic solutions of this theory. The duality symmetries
and the black hole solutions are reviewed in Sec. \ref{dsec}.

An instanton describing pair creation of charged black holes in
background fields in this theory was obtained in \cite{2u1}, and is
reviewed in Sec. \ref{Esol}. This instanton is obtained from a
generalisation of the Ernst solution of Einstein-Maxwell theory so
that the black holes have two gauge charges and there are two
corresponding background fields. The instanton is very similar to the
Ernst instanton, but the presence of two background fields introduces
some interesting complications. In particular, the black holes are not
spherically symmetric in the extremal limit in this case, unlike the
Einstein-Maxwell case \cite{dggh}.

The main aim in this paper is to calculate the pair creation rate
given by this instanton, which will allow us to extend the conclusions
of \cite{dggh,entarea} to this case. The amplitude for pair creation
in the instanton approximation to the path integral is given by
$e^{-I}$, where $I$ is the action of the instanton. The pair creation
rate will thus be given by $e^{-I_b}$, where $I_b$ is the action of
the ``bounce'', an instanton--anti-instanton pair.  Sec. \ref{so4sec}
is thus dedicated to the calculation of the action for the bounce. We
find that the pair creation of non-extreme black holes is enhanced
over that for extreme black holes by $e^{{\cal A}_{bh}/4}$, from which
we conclude that the non-extreme black holes have $e^{{\cal
    A}_{bh}/4}$ more states than the extreme ones. That is, we
conclude that the number of states is given by $e^{S_{bh}}$.  In the
point-particle limit, where the black holes are small on the scale set
by the acceleration, the pair creation rate reduces to the Schwinger
result. That is, to leading order, the pair creation rate for the
black holes is the same as that for particles of the same mass and
charges. In summary, the results of this calculation of the pair
creation rate are essentially those of the calculation for the
Einstein-Maxwell theory in \cite{dggh,entarea}; this might not seem
surprising, as the Einstein-Maxwell theory is included as a special
case, but there is much more freedom in the $U(1)^2$ theory, so it is
a non-trivial result.

\section{Properties of the theory}
\label{dsec}

In \cite{Kalloshcens}, two actions were given for a low-energy theory
with two $U(1)$ gauge fields and a dilaton,
\begin{equation} \label{Kaction1}
I_{SO(4)} = -\frac{1}{16\pi} \int_M ( R - 2 \partial^{\mu}\phi
\partial_{\mu} \phi -(e^{2\phi} \tilde{F}^2 +
e^{-2\phi} G^2)) -\frac{1}{
8\pi}\oint_{\partial M} (K-K_0)
\end{equation}
and
\begin{equation} \label{Kaction2}
I_{SU(4)} = -\frac{1}{16\pi} \int_M ( R - 2 \partial^{\mu}\phi
\partial_{\mu} \phi - e^{-2\phi}( F^2 +
 G^2)) -\frac{1}{
8\pi}\oint_{\partial M} (K-K_0).
\end{equation}
These can be regarded as arising from the $SO(4)$ and $SU(4)$ versions
of $N=4$ supergravity respectively \cite{cremmer}. If we take
\begin{equation} \label{su4toso4}
\tilde{F}_{\mu\nu}  =  \frac{1}{2} e^{-2\phi}
\epsilon_{\mu\nu\rho\sigma} F^{\rho\sigma},
\end{equation}
It is easy to see that (\ref{Kaction1}) and (\ref{Kaction2}) give the
same equations of motion, but the values of these two actions are
different. To calculate the pair creation rate, we need to know which
of these actions we should take.

If we consider the Einstein-Maxwell case, where $\tilde{F}_{\mu\nu} =
G_{\mu\nu} = \frac{1}{\sqrt{2}} {\cal F}_{\mu\nu}$ (say), and
$\phi=0$, (\ref{Kaction1}) reduces to the usual Einstein-Maxwell
action, with Maxwell field ${\cal F}_{\mu\nu}$, while in
(\ref{Kaction2}), the two gauge field terms cancel. I therefore think
that, since we use the Einstein-Maxwell action in the calculation of
the pair creation rate in the Einstein-Maxwell case, we should use
(\ref{Kaction1}) to calculate the pair creation rate in this case.  In
\cite{2u1}, where the pair creation instanton was obtained, the
solutions were written in terms of $F$ and $G$. Since I will use
(\ref{Kaction1}) to calculate the pair creation rate, I will instead
write them here in terms of $\tilde{F}$ and $G$.

One interesting feature of the $U(1)^2$ theory is that it has two
distinct duality symmetries. The equations of motion of this theory are
invariant under a duality transformation,
\begin{equation} \label{dualtransf}
F_{\mu\nu} \rightarrow \tilde{F}_{\mu\nu} \equiv  \frac{1}{2}  e^{-2\phi}
\epsilon_{\mu\nu\rho\sigma} F^{\rho\sigma},
\end{equation}
\begin{equation} \label{dualt2}
 G_{\mu\nu} \rightarrow
\tilde{G}_{\mu\nu} \equiv  \frac{1}{2}
e^{-2\phi} \epsilon_{\mu\nu\rho\sigma} G^{\rho\sigma},\, \phi
\rightarrow -\phi,
\end{equation}
which is analogous to the ordinary electric-magnetic duality
transformation of Einstein-Maxwell theory. The equations of motion and
the action (\ref{Kaction2}) are also invariant under the interchange
of the two gauge fields, $F_{\mu\nu} \leftrightarrow G_{\mu\nu}$. If
we combine these, we find that the equations of motion and
the action (\ref{Kaction1}) are invariant under the ``duality''
\begin{equation} \label{otherdual}
F_{\mu\nu} \to \tilde{G}_{\mu\nu}, G_{\mu\nu} \to \tilde{F}_{\mu\nu},
\phi \to -\phi.
\end{equation}
If we think of $\tilde{F}$ as the field variable rather than $F$, this
``duality'' just interchanges the two gauge fields and  reverses
the sign of the dilaton.

On the Einstein-Maxwell solutions, for which $\tilde{F}_{\mu\nu} =
G_{\mu\nu}$ and $\phi=0$, (\ref{otherdual}) is a trivial
transformation. In general, we will consider solutions for which the
transformation (\ref{otherdual}) just corresponds to an interchange of
the parameters of the solution. These solutions will be said to have a
manifest duality symmetry. The action (\ref{Kaction1}) is invariant
under this manifest duality symmetry.

The charged black hole solutions of the $U(1)^2$ theory are \cite{Gary}:
\begin{equation}
ds^2 = -\lambda dt^2 + \lambda^{-1} dr^2 + R^2 d\Omega, \label{Kbhole1}
\end{equation}
\begin{equation}
e^{2\phi} = e^{2\phi_0} \frac{r+\Sigma}{r-\Sigma}, \label{Kbhole2}
\end{equation}
\begin{equation}
\tilde{F} =Q e^{-\phi_0}\sin \theta d\theta \wedge d\varphi,\; G
=Pe^{\phi_0}\sin \theta d\theta \wedge d\varphi, \label{Kbhole3}
\end{equation}
where
\begin{equation} \label{Klambda}
\lambda = \frac{(r-r_+)(r-r_-)}{R^2}, \; R^2 = r^2 -\Sigma^2,
\end{equation}
and \cite{stringbh}
\begin{equation} \label{Kparam}
r_{\pm} = M \pm \sqrt{M^2+\Sigma^2-P^2-Q^2},\; \Sigma = \frac{P^2-
Q^2}{2M}.
\end{equation}
There is a curvature singularity at $r=|\Sigma|$. The physical degrees
of freedom are $P, Q, M$ and $\phi_0$; $M$ is the mass of the black
hole, and $e^{-\phi_0} Q$ and $e^{\phi_0} P$ are its gauge charges.
Note that both the gauge fields are magnetic, when we write the
solutions this way. One can also obtain a solution with two electric
fields, but I will restrict attention to the magnetic case.  We could
keep the asymptotic value of the dilaton $\phi_0$ as a free parameter,
but I will instead fix it by requiring that the dilaton match to an
appropriate background value at infinity. The solution has a manifest
duality symmetry, as the
solution is unchanged when
\begin{equation} \label{bhsymm}
 \tilde{F} \leftrightarrow G,\; \phi \leftrightarrow -\phi,
\end{equation}
and
\begin{equation}
  Q \leftrightarrow P,\; \Sigma \leftrightarrow -\Sigma,\; \phi_0
  \leftrightarrow - \phi_0.
\end{equation}

\section{The pair creation instanton}
\label{Esol}

The pair creation of black holes is described by an instanton, that
is, a solution of the classical equations of motion with Euclidean
signature, which acts as a saddle-point in the path integral. The
solution which gives the instanton in the $U(1)^2$ theory, which I
will refer to as the $U(1)^2$ Ernst solution, was obtained
in \cite{2u1}. It is a generalisation of the Ernst solution of
Einstein-Maxwell theory \cite{Ernst}. Like the Ernst solution, it
describes a pair of oppositely-charged black holes undergoing uniform
acceleration under the influence of background electromagnetic fields.
It asymptotically approaches an analogue of the Melvin solution
\cite{melvin}, which describes the background fields,  which I will refer
to as the $U(1)^2$ Melvin solution.

The $U(1)^2$ Melvin solution is
\begin{equation}
\label{dualMelvin} ds^2 = \Lambda \Psi[-dt^2+d\rho^2+dz^2] +
\frac{\rho^2 d\varphi^2}{\Lambda \Psi},
\end{equation}
\begin{equation} \label{dMgauge}
e^{-2\phi} = \frac{\Lambda}{\Psi},\, A_\varphi =
-\frac{\widehat{B}_M\rho^2}{2\Lambda},\,
B_\varphi = -\frac{\widehat{E}_M\rho^2}{2\Psi},
\end{equation}
\begin{equation}
G_{\mu\nu} = \partial_{[\mu} A_{\nu]},\, \tilde{F}_{\mu\nu} =
\partial_{[\mu} B_{\nu]},
\end{equation}
\begin{equation}
\Lambda = 1+ \frac{1}{2}  \widehat{B}_M^2 \rho^2,\, \Psi =
1+\frac{1}{2} \widehat{E}_M^2
\rho^2.
\end{equation}
This solution has a manifest duality symmetry under
\begin{equation}
 \tilde{F} \leftrightarrow G,\,
\phi \leftrightarrow  -\phi,
\mbox{ and } \widehat{B}_M \leftrightarrow \widehat{E}_M.
\end{equation}
It represents a pair of magnetic fields which are essentially uniform
near the axis $\rho=0$, with field strengths given by $\widehat{E}_M$
and $\widehat{B}_M$. The fields depart from uniformity away from
the axis because the field energy curves the spacetime. However, in
practice we cannot construct such strong fields, so the physically
interesting part of this solution is the region near the axis.

The $U(1)^2$ Ernst solution is
\begin{eqnarray}
ds^2 &=& \frac{\Lambda\Psi}{A^2(x-y)^2}[F(x)(G(y)dt^2-G^{-1}(y)
dy^2) \label{Ernst} \\
&& +F(y)G^{-1}(x) dx^2]+ \frac{F(y)G(x)}{\Lambda\Psi A^2(x-y)^2} d\varphi^2,
\nonumber
\end{eqnarray}
\begin{equation} \label{Ernstf}
e^{-2\phi} = e^{-2\phi_0} \frac{\Lambda}{\Psi} \left(\frac{1+\Sigma
A y}{1- \Sigma Ay}\right) \left(\frac{1-\Sigma Ax}{1+\Sigma A x}
\right),
\end{equation}
\begin{equation}
A_{\varphi} = -\frac{e^{\phi_0}}{B\Lambda}\left(1+\frac{B\beta
x}{1-
\Sigma Ax}\right)+k,
\end{equation}
\begin{equation}
B_{\varphi} = -\frac{e^{-\phi_0}}{E \Psi}
\left(1+ \frac{E \alpha x}{1 + \Sigma Ax}\right)+k',
\end{equation}
\begin{equation}
G_{\mu\nu} = \partial_{[\mu} A_{\nu]},\,
 \tilde{F}_{\mu\nu} = \partial_{[\mu} B_{\nu]},
\end{equation}
where
\begin{eqnarray}
\Lambda &=& \left(1+\frac{B\beta x}{1-\Sigma Ax}\right)^2  \\
&& +\frac{B^2(1-
x^2-r_+ A x^3)(1+r_- A x)(1-\Sigma Ay)^2}{2A^2(x-y)^2(1-\Sigma A
x)^2}, \nonumber
\end{eqnarray}
\begin{eqnarray}
\Psi &=& \left(1+\frac{E\alpha x}{1+\Sigma Ax}\right)^2 \\
&&+ \frac{E^2(1-
x^2-r_+ A x^3)(1+r_- A x)(1+\Sigma Ay)^2}{2A^2 (x-y)^2(1+\Sigma A
x)^2}, \nonumber
\end{eqnarray}
\begin{equation} \label{Cmfnsf}
F(\xi) = 1-\Sigma^2 A^2 \xi^2,
\end{equation}
\begin{equation} \label{Cmfnsg}
 G(\xi) = \frac{(1-\xi^2-r_+ A
\xi^3)(1+r_- A\xi)}{(1- \Sigma^2 A^2 \xi^2)},
\end{equation}
and
\begin{eqnarray} \label{alpha}
\alpha^2 &=& \frac{1}{2}  (r_+ -\Sigma)(r_- -\Sigma) +\frac{1}{2}  A^2
\Sigma^3 (r_-  -\Sigma)\\
&=& Q^2
+ \frac{1}{2}  A^2 \Sigma^3 (r_- -
\Sigma),  \nonumber
\end{eqnarray}
\begin{eqnarray} \label{beta}
\beta^2 &=& \frac{1}{2}  (r_+ + \Sigma)(r_- + \Sigma) - \frac{1}{2}
A^2 \Sigma^3 (r_-  + \Sigma) \\
&=& P^2 -\frac{1}{2}  A^2 \Sigma^3 (r_- +
\Sigma). \nonumber
\end{eqnarray}
As we will see below, this solution represents a pair of
oppositely-charged black holes accelerating away from each other in a
background field, although the coordinate system used here only
includes one of the black holes. The black holes carry two magnetic
gauge charges, and the background consists of two magnetic fields,
which reduce to the fields in the $U(1)^2$ Melvin solution if we go to
infinity along the axis of symmetry. The constants $\phi_0, k$, and
$k'$ will be chosen so that the solution at infinity agrees with
(\ref{dualMelvin}).

For $r_+ A < 2/(3\sqrt{3})$, the function $G(\xi)$ has four real
roots, which I denote in ascending order by $\xi_1,\, \xi_2,\,
\xi_3,\, \xi_4$. It is convenient to define another function $H(\xi) =
G(\xi) F(\xi)$, so that I may write
\begin{equation}
H(\xi) = -(r_+A)(r_-A)(\xi-\xi_1)(\xi-\xi_2)(\xi-\xi_3)(\xi-\xi_4).
\end{equation}
I restrict the parameters so that $\xi_1 = -1/r_- A$ and $\xi_1 \leq
\xi_2 \leq \xi_3 < \xi_4$.  The surface $y = \xi_0 \equiv -1/|\Sigma|
A$ is singular; this is the singular surface inside the black hole,
that is, the singular surface at $r=|\Sigma|$ in the black hole
solutions (\ref{Kbhole1}). As $r_- \geq |\Sigma|,\, \xi_1 \geq \xi_0$.
The surfaces $y=\xi_1,\, y=\xi_2$ are the inner and outer black hole
horizons, and $y=\xi_3$ is the acceleration horizon for an observer
comoving with the black hole. The coordinates $(x,\varphi)$ are
angular coordinates which cover two-spheres around the black hole,
except when $y=\xi_3$. So that the metric has the appropriate
signature, $x$ is restricted to the range $\xi_3 \leq x \leq \xi_4$ in
which $G(x)$ is positive.  At $x=\{\xi_3 ,\, \xi_4\}$, the norm of
$\partial/\partial \varphi$ vanishes, so these points are interpreted
as the poles of the two-spheres; that is, the axis of symmetry is
$x=\xi_3,\xi_4$, with $x=\xi_3$ pointing at infinity, and $x=\xi_4$
pointing at the other black hole. There is a divergence in the metric
at $x=y$, which is interpreted as the point at infinity, so $y$ is
restricted to the range $\xi_0 < y < x$.  Spatial infinity is reached
only along the axis, that is, when $y=x=\xi_3$, and null or timelike
infinity when $y=x\neq \xi_3$ \cite{ashtekar}.

This solution has a manifest duality symmetry under
\begin{equation} \label {Ernstsym}
  \tilde{F} \leftrightarrow G,\,
  \phi \leftrightarrow -\phi,
\end{equation}
and
\begin{equation}
   Q \leftrightarrow P,\, \Sigma \leftrightarrow -\Sigma,\, B
  \leftrightarrow E,\, k \leftrightarrow k',\, \phi_0 \leftrightarrow
-\phi_0.
\end{equation}

As in the Ernst solution \cite{Ernst}, the background fields provide the
force necessary to accelerate the black holes. To eliminate the nodal
singularities in this metric at $x=\xi_3$ and $x=\xi_4$
simultaneously, $A$ must be chosen so that\footnote{Note that
$\Lambda(\xi_i) \equiv \Lambda(x=\xi_i)$ and $\Psi(\xi_i) \equiv
\Psi(x=\xi_i)$ are constants.}
\begin{equation} \label{constraint2}
G'(\xi_3) \Lambda(\xi_4) \Psi(\xi_4) = -G'(\xi_4)
\Lambda(\xi_3) \Psi(\xi_3)
\end{equation}
 and we must take $\Delta \varphi=4\pi L^2 / G'(\xi_3)$, where I have
introduced $L^2 = \Lambda(\xi_3) \Psi(\xi_3)$. In the limit $r_+ A \ll
1$, (\ref{constraint2}) reduces to Newton's law, $MA \approx BP + EQ$,
and in general it determines the acceleration of the black holes in
terms of the other parameters.

If I set $r_+ = r_- =0$, (\ref{Ernst}) becomes
\begin{eqnarray} \label{accmel}
ds^2 &= & \frac{\Lambda\Psi}{A^2(x-y)^2}[(1-y^2)dt^2-(1-
y^2)^{-1} dy^2\\
&& +(1-x^2)^{-1} dx^2]
+ \frac{1-x^2}{\Lambda\Psi A^2(x-y)^2} d\varphi^2, \nonumber
\end{eqnarray}
where
\begin{equation}
\Lambda = 1+ \frac{1}{2}  B^2 \frac{1-x^2}{A^2(x-y)^2},
\end{equation}
and
\begin{equation}
\Psi =  1+ \frac{1}{2}  E^2 \frac{1-x^2}{A^2(x-y)^2}.
\end{equation}
This is just the $U(1)^2$ Melvin solution (\ref{dualMelvin}) in
non-standard coordinates \cite{2u1}. That is, the $U(1)^2$ Melvin
solution is a special case of the $U(1)^2$ Ernst solution, where the
black hole parameters are set to zero.

The $U(1)^2$ Ernst solution (\ref{Ernst}) also approaches
(\ref{dualMelvin}) at large spacelike distances, that is, when we go
to infinity along the axis. Spatial infinity corresponds to $x, y
\rightarrow \xi_3$, and in this limit it is convenient to use the
change of coordinates given in \cite{dggh},
\begin{equation} \label{inftransf}
x - \xi_3 = \frac{4 F(\xi_3)
L^2}{G'(\xi_3)A^2}\frac{\rho^2}{(\rho^2+\zeta^2)^2},
\end{equation}
\begin{equation}
\xi_3 -y =
\frac{4 F(\xi_3)L^2}{G'(\xi_3) A^2}
\frac{\zeta^2}{(\rho^2+\zeta^2)^2},
\end{equation}
\begin{equation}
t= \frac{2 \eta}{G'(\xi_3)},\, \varphi = \frac{2L^2
\tilde{\varphi}}{G'(\xi_3)}.
\end{equation}
For large $\rho^2+\zeta^2$, the $U(1)^2$ Ernst solution in these
coordinates reduces to
\begin{equation}
ds^2 \rightarrow \tilde{\Lambda}\tilde{\Psi} (-\zeta^2 d\eta^2 +
d\zeta^2 + d\rho^2) + \frac{\rho^2
d\tilde{\varphi}^2}{\tilde{\Lambda}\tilde{\Psi}},
\end{equation}
where
\begin{equation}
\tilde{\Lambda} = (1+\frac{1}{2}  \widehat{B}_E^2 \rho^2) \mbox{ with
} \widehat{B}_E^2  =\frac{B^2  G'^2(\xi_3)}{4L^2\Lambda(\xi_3)},
\end{equation}
and
\begin{equation}
\tilde{\Psi} = (1+\frac{1}{2}  \widehat{E}_E^2 \rho^2) \mbox{ with
} \widehat{E}_E^2 =  \frac{E^2 G'^2(\xi_3)}{4L^2\Psi(\xi_3)}.
\end{equation}
If I now set $\hat{t} = \zeta \sinh \eta, z = \zeta \cosh \eta$, we
once again regain (\ref{dualMelvin}). For large $\rho^2+\zeta^2$, the
dilaton and gauge fields tend to
\begin{equation}
e^{-2\phi} \rightarrow  e^{-2\phi_0} \frac{\Lambda(\xi_3)}{\Psi(\xi_3)}
\frac{\tilde{\Lambda}}{\tilde{\Psi}},
\end{equation}
\begin{equation}
A_{\tilde{\varphi}} \rightarrow  e^{\phi_0}
\frac{\Psi(\xi_3)^{1/2}}{\Lambda(\xi_3)^{1/2}}
\frac{\hat{B}\rho^2}{ 2\tilde{\Lambda}},\, B_{\tilde{\varphi}}
\rightarrow  e^{-\phi_0}
\frac{\Lambda(\xi_3)^{1/2}}{\Psi(\xi_3)^{1/2}} \frac{\hat{E}
\rho^2}{2\tilde{\Psi}},
\end{equation}
so if I set $e^{2\phi_0}=\Lambda(\xi_3)/\Psi(\xi_3)$, I recover
(\ref{dMgauge}) in this limit. I will take this to define $\phi_0$ in
general. Thus, I recover the $U(1)^2$ Melvin solution at large
spacelike distances, and this allows me to identify the physical
strength of the background fields in the Ernst solution as
$\widehat{E}_E$ and $\widehat{B}_E$. We can also calculate the
physical charges on the black hole by integrating the field tensors
over two-spheres surrounding the black holes. We find
\begin{equation}
\widehat{P} = \frac{1}{4\pi} \int G = \frac{\Lambda(\xi_3)
  \Psi(\xi_3)^{3/2}}{ G'(\xi_3) \Lambda(\xi_4)^{1/2}} \frac{\beta
  (\xi_4 - \xi_3)}{ (1-\Sigma A \xi_4) (1 - \Sigma A \xi_3)}
  \label{physP}
\end{equation}
and
\begin{equation}
\widehat{Q} = \frac{1}{4\pi} \int \tilde{F} = \frac{\Psi(\xi_3)
  \Lambda(\xi_3)^{3/2}}{ G'(\xi_3) \Psi(\xi_4)^{1/2}} \frac{\alpha
    (\xi_4 -\xi_3)}{ (1+\Sigma A \xi_4) (1+\Sigma A \xi_3)},
  \label{physQ}
\end{equation}
where $\alpha$ and $\beta$ are given by (\ref{alpha},\ref{beta}).

The solution (\ref{Ernst}) describes two black holes accelerating away
from each other, propelled by the background fields.  Now we take the
Euclidean section obtained by taking $\tau = it$ in (\ref{Ernst}).
Half the Euclidean section gives an instanton describing black hole
pair production \cite{gwg,garstrom}. There are three possible
instantons: one describing pair production of non-extreme black holes,
one describing pair production of extreme black holes, with
$\xi_1=\xi_2$, and another special case when $\xi_2 = \xi_3$. We will
not consider this last here, as it does not describe black hole pair
production (see \cite{robb} for more details of this case).

Let us first consider the non-extreme or wormhole instantons, {\it
  i.e.}, $\xi_1<\xi_2< \xi_3$.  In the Euclidean section, we must
restrict $y$ to $\xi_2 \leq y \leq \xi_3$ to obtain a positive definite
metric, and $(y,\tau)$ are now also coordinates on a two-sphere,
except when $x=\xi_3$. We must impose another
condition on the parameters to eliminate the possible conical
singularities at the black hole horizon $y=\xi_2$ and the acceleration
horizon $y=\xi_3$ simultaneously. Namely, the period of $\tau$ must be
taken to be $\Delta \tau = 4\pi/|G'(\xi_2)|$, and we must set
\begin{equation}
|G'(\xi_2)| = |G'(\xi_3)|,
\end{equation}
where $G(\xi)$ is given by (\ref{Cmfnsg}).  This condition is
satisfied by setting
\begin{equation}
\left(\frac{\xi_2^2-\xi_0^2}{\xi_3^2-\xi_0^2}\right) \left(
\frac{\xi_3-\xi_1}{\xi_2-\xi_1} \right)= \frac{\xi_4-\xi_2}{\xi_4 -
\xi_3}. \label{instcond}
\end{equation}
This condition provides a further restriction on the black hole
parameters, which may be thought of as determining the mass of the
black hole in terms of its charges. More precisely, we can solve it
for $r_-A$ in terms of $r_+A$ and $\Sigma A$. The whole Euclidean
section is a bounce, that is, an instanton--anti-instanton pair joined
along a spacelike slice. The topology of the bounce is $S^2 \times S^2
- \{pt\}$, where the removed point is $x=y=\xi_3$.

For the extremal instantons, when $\xi_1 = \xi_2$, we must take
$\Delta \tau = 4\pi/|G'(\xi_3)|$ to ensure regularity at the
acceleration horizon. The black hole event horizon is at infinite
distance in all spatial directions, so we do not have to worry about a
conical singularity there.  The range of $y$ in the Euclidean section
is now $\xi_2 < y \leq \xi_3$, so that $(y,\tau)$ are now polar
coordinates on an $R^2$, except when $x=\xi_3$. The extremal bounce
has topology $S^2 \times R^2 - \{pt\}$, and the instanton can be
interpreted as creating a pair of extremal black holes, with
infinitely long throats.

However, I have found that, unlike the case with one $U(1)$ gauge
field \cite{dggh}, the extremal solutions do not become spherically
symmetric near the event horizon, and therefore do not approach the
static black hole solutions at this internal infinity. This can be
most easily seen by computing the intrinsic curvature scalar $^2 R$
for the black hole horizon itself when the black holes are extremal,
and calculating its numerical values at some typical horizon
positions. I will omit the rather unilluminating formula for $^2 R$,
and simply state that one finds that the curvature is larger at the
poles than at the equator of the two-sphere. Since the horizon is not
a round two-sphere, the solution cannot be spherically symmetric. The
point of this is that, unlike the case with one gauge field, even the
extremal black holes are accelerating in some sense. It would be
interesting to see if this could be extended to a Kaluza-Klein theory
with two gauge fields, as in the usual Kaluza-Klein theory there is a
well-defined sense in which the extremal black holes move on
geodesics, and are thus not accelerating.

Another difference that it is worth highlighting is that, even once
the no-strut condition (\ref{constraint2}) and either (\ref{instcond})
or $\xi_1 =\xi_2$ have been satisfied, there are still four parameters
in the solution, the two charges $\widehat{Q}$ and $\widehat{P}$ of
the black hole and the background field strengths $\widehat{B}_E$ and
$\widehat{E}_E$. This means we have a lot more freedom than in the
Einstein-Maxwell case, where we only had two parameters once the
regularity constraints were satisfied. In particular, if $\widehat{Q}$
and $\widehat{P}$ have opposite signs, it is possible to take large
values of $\widehat{B}_E$ and $\widehat{E}_E$ without producing very
large accelerations. This implies that, unlike the case with one
$U(1)$ gauge field \cite{dggh}, there does not seem to be any
universal bound on $\widehat{Q} \widehat{E}_E$ or $\widehat{P}
\widehat{B}_E$.

\section{The pair creation rate}
\label{so4sec}

Having described the pair creation instanton, I now turn to the
calculation of the pair creation rate. The principal results are that
the pair creation rate for non-extreme black holes is enhanced over
that for extreme black holes by $e^{{\cal A}_{bh}/4}$ (as in the
Einstein-Maxwell case \cite{dggh}), the pair creation rate is always
suppressed, and it reduces to the Schwinger result in the limit of
small black holes. The $U(1)^2$ Ernst metric reduces to the Ernst
metric when $\phi=0$, and to the dilaton Ernst metric when either
$\tilde{F}$ or $G$ vanishes, and so I can check the calculation by
showing that it agrees with the results of \cite{entarea,dggh} in
these cases.

The amplitude for pair creation in a background field is given by the
path integral
\begin{equation}
\Psi = \int d[g] d[A] d[B] e^{-I},
  \label{pint}
\end{equation}
where the action $I$ in the path integral is the action
(\ref{Kaction}), and the integral is over all metrics and gauge
fields which interpolate between the background fields at infinity and
a spacelike slice which contains the pair of black holes. If there is
an appropriate instanton, we assume that $\Psi$ will be approximately
$\Psi \approx e^{-I}$, where $I$ is now the action of the
instanton. The pair creation rate $\Gamma$ is given by the modulus
squared of this amplitude, so it will be approximately $\Gamma \approx
e^{-I_b}$, where $I_b$ is the action of the bounce. For the pair
creation of black holes, the Euclidean sections of the solutions
discussed in Sec. \ref{Esol} are the bounces, so the calculation
of the pair creation rates reduces to the problem of the calculation
of the actions of these bounces.

The simplest way to evaluate the action is by a Hamiltonian
decomposition, following the techniques given in \cite{haho}. Since
the solutions we are interested in are stationary, if the Euclidean
section was of the form $\Sigma \times S^1$, where the $S^1$ factor
represents the time direction, the action would just be given by $I =
\beta H$, where $H$ is the Hamiltonian and $\beta = \Delta \tau$ is
the period in imaginary time. However, the time-translation Killing
vector has fixed points at the black hole event horizon and the
acceleration horizon, so by doing this we have neglected a
contribution from a neighbourhood of each horizon. Including the
contributions from these corners, the
total Euclidean action is (in the non-extreme case)
\begin{equation} \label{impeq}
 I = \beta H -\frac{1}{ 4} (\Delta {\cal A} + {\cal A}_{bh}),
\end{equation}
where ${\cal A}_{bh}$ is the area of the black hole horizon, and
$\Delta {\cal A}$ is the difference in area of the acceleration
horizon between the solution and the background \cite{haho,entarea}.
In the extreme case, the term proportional to ${\cal A}_{bh}$ is
absent, as the black hole event horizon is not part of the Euclidean
section.  The Hamiltonian $H$, which is only defined
with respect to the background spacetime, can be expressed as \cite{haho}
\begin{equation} \label{hamil}
 H = \int_{\Sigma}
   N{\cal H} - \frac{1}{ 8\pi} \int_{S^\infty} N({}^2 K - {}^2K_0),
\end{equation}
where $N$ is the lapse, ${\cal H}$ is the Hamiltonian constraint,
${}^2 K$ is the trace of the two dimensional extrinsic curvature of
the boundary near infinity, and ${}^2 K_0$ is the analogous quantity
for the background spacetime. On solutions, the constraint vanishes,
and so the only non-zero contribution comes from the gravitational
surface term.

To calculate this surface term, we need to introduce a boundary near
infinity, and calculate its extrinsic curvature in the instanton
and the background solution. To ensure that the boundary used in both
calculations is the same, I need to match the intrinsic features of
the boundary; that is, the induced metric, the gauge field, and the
value of the dilaton on the boundary.

I take the boundary in the $U(1)^2$ Ernst solution to be
\begin{equation} \label{bern}
 x = \xi_3 + \epsilon_E \chi, \ \  y = \xi_3 + \epsilon_E
(\chi-1),
\end{equation}
where $0 \leq \chi \leq 1$, and make the coordinate transformations
\begin{equation} \label{changei}
\varphi = \frac{ 2L^2 }{
G'(\xi_3)} \varphi',\ t = \frac{2 }{ G'(\xi_3)} t',
\end{equation}
and I assume that the boundary in the $U(1)^2$ Melvin
solution lies at
\begin{equation}  \label{melbdry}
  x = -1 + \epsilon_M \chi[1 + \epsilon_E f(\chi)],
\end{equation}
\begin{equation}
y = -1 +\epsilon_M (\chi-1) [1+\epsilon_E g(\chi)]
\end{equation}
in the accelerated coordinate system (\ref{accmel}).  Other choices
for the boundary in the $U(1)^2$ Melvin solution may be possible, but
this is the only choice that I have been able to explicitly carry out.
We evaluate all quantities to second nontrivial order in $\epsilon_E$,
as higher-order terms will not affect the result in the limit
$\epsilon_E \to 0$. For the $U(1)^2$ Ernst metric, the induced
metric on the boundary is
\begin{eqnarray}
  {}^{(2)} ds^2 &=& \frac{ L^2 F(\xi_3) }{ A^2 \epsilon_E G'(\xi_3)}
  \left\{ - \frac{ \lambda \psi d \chi^2 }{ \chi (\chi-1)} \left[
    1+\epsilon_E (2 \chi -1) \frac{F'(\xi_3) }{ F(\xi_3)} \right]
  \right. \\ \nonumber &&+ \left.  \frac{4 \chi }{\lambda \psi}
  \left[1 + \epsilon_E \chi \frac{H''(\xi_3) }{ 2 H'(\xi_3)}
    -\epsilon_E \frac{F'(\xi_3) }{ F(\xi_3)} \right] d\varphi'^2
  \right\}, \label{emetric}
\end{eqnarray}
where
\begin{equation} \label{lambda}
\lambda = 1 + \frac{ 2 L^2 \widehat{B}_E^2 F(\xi_3)\chi}{ A^2 \epsilon_E
G'(\xi_3)} \left( 1 + \frac{1}{2} \epsilon_E \chi
\frac{H''(\xi_3)}{H'(\xi_3)} + \frac{2 \Sigma A \epsilon_E}{1-\Sigma A
\xi_3} \right)
\end{equation}
and
\begin{equation} \label{psi}
\psi  = 1 + \frac{ 2 L^2 \widehat{E}_E^2 F(\xi_3) \chi}{ A^2 \epsilon_E
G'(\xi_3)} \left( 1 + \frac{1}{2} \epsilon_E \chi
\frac{H''(\xi_3)}{H'(\xi_3)} - \frac{2 \Sigma A \epsilon_E}{1 + \Sigma
A \xi_3} \right).
\end{equation}
The gauge potentials on the boundary for the $U(1)^2$ Ernst solution are
\begin{equation} \label{fgauge}
  A_{\varphi'} = \frac{ L e^{\phi_0}}{\Lambda(\xi_3) \widehat{B}_E}
  \left[ 1 - \frac{A^2 \epsilon_E G'(\xi_3)}{2 L^2 F(\xi_3)
      \widehat{B}_E^2 \chi} \right]
\end{equation}
and
\begin{equation} \label{ggauge}
B_{\varphi'} = \frac{ L e^{-\phi_0}}{\Psi(\xi_3) \widehat{E}_E} \left[
1 - \frac{A^2 \epsilon_E G'(\xi_3)}{2 L^2 F(\xi_3) \widehat{E}_E^2
\chi} \right].
\end{equation}
The dilaton at the boundary is
\begin{equation} \label{dilb}
e^{-2\phi} = e^{-2\phi_0} \frac{\Lambda(\xi_3) \lambda}{\Psi(\xi_3)
\psi} \left( 1 -\frac{2 \Sigma A \epsilon_E}{1 - \Sigma^2 A^2 \xi_3^2}
\right).
\end{equation}

For the $U(1)^2$ Melvin solution, the induced metric on the boundary is
\begin{eqnarray} \label{bmetM}
  {}^{(2)} ds^2 &=& \frac{-\Lambda \Psi }{ 2 \chi (\chi-1) \bar{A}^2
    \epsilon_M} \left\{ 1 - \epsilon_E (\chi-1) f(\chi) + \epsilon_E
  \chi g(\chi) \right. \\ \nonumber && -2 \epsilon_E \chi (\chi-1)
  [f'(\chi) -g'(\chi)] - \left. 2 \epsilon_E [\chi f(\chi) - (\chi-1)
  g(\chi)] \right\} d\chi^2
\\ \nonumber &&
+\frac{2 \chi }{\Lambda \Psi  \bar{A}^2 \epsilon_M}\left\{ 1
-\frac{1}{2} \epsilon_M \chi
+ \epsilon_E f(\chi) -2 \epsilon_E [\chi f(\chi) - (\chi-1) g(\chi)]
\right\} d\varphi^2,
\end{eqnarray}
where
\begin{eqnarray} \label{lambdaM}
\Lambda &=& 1+\frac{ \widehat{B}_M^2 \chi }{ \bar{A}^2
\epsilon_M} \left\{ 1 - \frac{1}{ 2} \epsilon_M \chi + \epsilon_E
f(\chi) \right. \\ \nonumber
&&- \left. 2 \epsilon_E [\chi f(\chi) - (\chi-1) g(\chi)]\right\}
\end{eqnarray}
and
\begin{eqnarray} \label{psiM}
\Psi &=& 1+\frac{ \widehat{E}_M^2 \chi }{\bar{A}^2
\epsilon_M} \left\{ 1 - \frac{1}{ 2} \epsilon_M \chi + \epsilon_E
f(\chi) \right. \\ \nonumber
&&- \left. 2 \epsilon_E [\chi f(\chi) - (\chi-1) g(\chi)]\right\} .
\end{eqnarray}
The gauge potentials on the boundary in $U(1)^2$ Melvin are
\begin{equation} \label{mgauge}
A_{\varphi} = \frac{1}{ \widehat{B}_M} \left[
1 - \frac{ \bar{A}^2 \epsilon_M}{\widehat{B}_M^2 \chi} \right]
\end{equation}
and
\begin{equation} \label{mgauge2}
B_{\varphi} = \frac{1}{ \widehat{E}_M} \left[
1 - \frac{ \bar{A}^2 \epsilon_M}{\widehat{E}_M^2 \chi} \right],
\end{equation}
and the dilaton at the boundary in $U(1)^2$ Melvin is
\begin{equation} \label{mdil}
e^{-2 \phi} = \frac{\Lambda}{\Psi},
\end{equation}
where $\Lambda$ is given by (\ref{lambdaM}) and $\Psi$ is given by
(\ref{psiM}).

I fix the remaining coordinate freedom by taking
\begin{equation} \label{abar2}
\bar{A}^2 = -\frac{G'(\xi_3) }{ 2 L^2 F(\xi_3)}\frac{H'(\xi_3)
}{ H''(\xi_3)}A^2,
\end{equation}
and write
\begin{equation} \label{expang}
 e^{\phi_0} = \frac{\Lambda(\xi_3)^{1/2}}{\Psi(\xi_3)^{1/2}} \left(1 -
\gamma  \epsilon_E
\right),  \widehat{B}_M = \widehat{B}_E \left(1+\alpha \epsilon_E
\right), \widehat{E}_M = \widehat{E}_E \left(1 + \beta \epsilon_E \right).
\end{equation}
I then find that the intrinsic metric, gauge potentials and dilaton
on the boundary can all be matched by taking
\begin{equation} \label{subl1}
\epsilon_M  = -\frac{H''(\xi_3)}{H'(\xi_3)} \epsilon_E,
\end{equation}
\begin{equation} \label{subl2}
f(\chi) =  \frac{F'(\xi_3) }{ F(\xi_3)} (4\chi -3),\ \ g(\chi) =
\frac{F'(\xi_3) }{ F(\xi_3)} (4\chi-1),
\end{equation}
and
\begin{equation} \label{subl3}
\gamma = \alpha = -\beta = \frac{\Sigma A}{1-\Sigma^2 A^2 \xi_3^2}.
\end{equation}

Note that the lapse function is also matched by these
conditions. For the $U(1)^2$ Ernst metric, the lapse function at the
boundary is given by
\begin{equation}
N = \left[\frac{4 L^2 F(\xi_3)(1-\chi)\lambda \psi}{ A^2 \epsilon_E
G'(\xi_3)}\right]^{\frac{1}{2}} \left[ 1 +
\frac{1}{4} \epsilon_E (\chi-1)
\frac{H''(\xi_3) }{ H'(\xi_3)} + \frac{1}{2} \epsilon_E
\frac{F'(\xi_3) }{ F(\xi_3)} \right], \label{lapseE}
\end{equation}
While the lapse function for the $U(1)^2$ Melvin metric is
\begin{eqnarray} \label{lapseM}
N &=& \left[\frac{2(1-\chi)\Lambda \Psi}{ \bar{A}^2
\epsilon_M}\right]^{\frac{1}{2}}
 \left\{ 1 - \frac{1}{4} \epsilon_M (\chi-1)+ \frac{1}{2} \epsilon_E
g(\chi) \right. \\ \nonumber && \left. - \epsilon_E
[\chi f(\chi) - (\chi-1) g(\chi)] \phantom{\frac{1}{2}} \right\},
\end{eqnarray}
so we see that the matching conditions (\ref{abar2}-\ref{subl3}) make
(\ref{lapseE}) and (\ref{lapseM}) equal as well.

The extrinsic curvature of this boundary embedded in the $U(1)^2$ Ernst
solution is
\begin{eqnarray} \label{excurv2}
{}^2 K &=& \frac{A \epsilon_E^{1/2} G'(\xi_3)^{1/2} }{ L
F(\xi_3)^{1/2} \lambda \psi} \left[ 1+ \frac{1}{4}
\epsilon_E \frac{H''(\xi_3) }{ H'(\xi_3)} (4 \chi -3) \right. \\
\nonumber
&&- \left.\frac{1 }{2}\epsilon_E \frac{F'(\xi_3) }{ F(\xi_3)} (4 \chi
-3) \right],
\end{eqnarray}
while the extrinsic curvature of the boundary embedded in the $U(1)^2$
Melvin solution is
\begin{eqnarray} \label{excurvM2}
{}^2 K_0 &=& \frac{\bar{A} \epsilon_M^{1/2} \sqrt{2} }{
\Lambda \Psi} \left[ 1 -\frac{1}{4} \epsilon_M (4\chi-3)
\right. \\ \nonumber &&- \left.
\frac{1 }{ 2} \epsilon_E \frac{F'(\xi_3) }{ F(\xi_3)} (24\chi
-13) \right].
\end{eqnarray}
Using the matching conditions (\ref{abar2}-\ref{subl3}), one may
now evaluate
\begin{equation} \label{ecdiff}
{}^2 K - {}^2 K_0 = \frac{5 A \epsilon_E^{3/2} G'(\xi_3)^{1/2} }{ L
F(\xi_3)^{1/2} \lambda \psi} \frac{F'(\xi_3) }{ F(\xi_3)}
(2\chi -1).
\end{equation}
Therefore, taking the limit $\epsilon_E \rightarrow 0$,
the Hamiltonian is
\begin{equation} \label{hameval}
 H_E = -\frac{1}{4} \int_0^1 d \chi N \sqrt{h} (^2 K - {}^2
K_0) = - \frac{5 L^2 F'(\xi_3) }{ A^2 G'(\xi_3)} \int_0^1
d\chi (2 \chi -1) =0.
\end{equation}
The action is thus given by
\begin{equation}
I = -\frac{1}{4} (\Delta {\cal A} + {\cal A}_{bh} )
  \label{keyeq}
\end{equation}
when the black holes are non-extremal, and by
\begin{equation}
I =  -\frac{1}{4} \Delta {\cal A}
  \label{keyeqq}
\end{equation}
if the black holes are extremal. Note that the action in the
non-extreme case is less than the action in the extreme case by
$-\frac{1}{4} {\cal A}_{bh}$, and thus the pair creation rate for
non-extreme black holes in enhanced by $e^{{\cal A}_{bh}/4}$ over that
for extreme black holes. A natural interpretation of this result is
that the non-extreme black holes have $e^{{\cal A}_{bh}/4}$ more
states than the extreme ones, as in the Einstein-Maxwell case
\cite{entarea}. As the difference in entropy between the non-extreme
and extreme solutions is also $\frac{1}{4} {\cal A}_{bh}$
\cite{entarea}, this suggests that the entropy is a reliable guide to
the number of states, that is, the number of states of a black hole
$\sim e^{S_{bh}}$.

I now proceed to calculate the right hand side of (\ref{keyeq}) and
(\ref{keyeqq}). The area of the black hole horizon is
\begin{equation} \label{bhhoriz}
{\cal A}_{bh}  = \int_{y=\xi_2} \sqrt{g_{xx} g_{\varphi\varphi}} dx
d\varphi  = \frac{4\pi F(\xi_2) L^2}{A^2 G'(\xi_3)}
\frac{(\xi_4-\xi_3)}{(\xi_3-\xi_2)(\xi_4-\xi_2)}.
\end{equation}
Again, in the calculation of the difference in area of the
acceleration horizon, I need to introduce a boundary, in this case a
circle, at large distances, and match the intrinsic features of this
boundary.  The area of the acceleration horizon in the $U(1)^2$ Ernst
spacetime up to a large circle at $x = \xi_3 + \epsilon_E$ is
\begin{equation} \label{eaccel}
{\cal A}_E  = \int_{y=\xi_3} \sqrt{g_{xx} g_{\varphi\varphi}} dx
d\varphi  = - \frac{4 \pi L^2 F(\xi_3)}{A^2 G'(\xi_3) (\xi_4 - \xi_3)} + \pi
\rho_E^2,
\end{equation}
where $\rho_E^2 = 4 F(\xi_3) L^2/[G'(\xi_3) A^2 \epsilon_E]$. The area
of the acceleration horizon in the $U(1)^2$ Melvin spacetime inside a
circle at $\rho = \rho_M$ is ${\cal A}_M = \pi \rho_M^2$.

Now I need to match the proper length of the boundary, the integral of
both gauge potentials around the boundary, and the value of $\phi$ at
the boundary. The proper length of the boundary in the $U(1)^2$ Ernst
solution is
\begin{eqnarray} \label{lengE}
  l_E &= &\frac{4\pi}{\widehat{E}_E \widehat{B}_E \rho_E} \left[ 1 -
  \frac{F(\xi_3) L^2}{G'(\xi_3) A^2} \frac{H''(\xi_3) }{H'(\xi_3)}
  \frac{1}{\rho_E^2} + \frac{2 F(\xi_3) L^2}{ G'(\xi_3) A^2}
  \frac{F'(\xi_3)}{F(\xi_3)} \frac{1}{\rho_E^2} \right. \\ \nonumber
  && \left. - \frac{1}{\widehat{E}_E^2 \rho_E^2} -
  \frac{1}{\widehat{B}_E^2 \rho_E^2} \right],
\end{eqnarray}
while the proper length of the boundary in the $U(1)^2$ Melvin solution is
\begin{equation} \label{lengM}
l_M = \frac{4 \pi}{\widehat{E}_M \widehat{B}_M \rho_M} \left( 1
-\frac{1}{\widehat{E}_M^2 \rho_M^2} - \frac{1}{ \widehat{B}_M^2
\rho_M^2} \right).
\end{equation}
The integral of the gauge potentials around the boundary are, in the
$U(1)^2$ Ernst solution,
\begin{equation} \label{pot1E}
\oint A_\varphi d\varphi = \frac{2 \pi}{\widehat{B}_E} \frac{
e^{\phi_0} \Psi(\xi_3)^{1/2}}{\Lambda(\xi_3)^{1/2}} \left( 1 -
\frac{2}{\widehat{B}_E^2 \rho_E^2} \right)
\end{equation}
and
\begin{equation} \label{pot2E}
\oint B_\varphi d\varphi = \frac{2 \pi}{\widehat{E}_E} \frac{
e^{-\phi_0} \Lambda(\xi_3)^{1/2}}{\Psi(\xi_3)^{1/2}} \left( 1 -
\frac{2}{\widehat{E}_E^2 \rho_E^2} \right).
\end{equation}
while in the $U(1)^2$ Melvin solution, they are
\begin{equation} \label{pot1M}
\oint A_\varphi d\varphi = \frac{2 \pi}{\widehat{B}_M}  \left( 1 -
\frac{2}{\widehat{B}_M^2 \rho_M^2} \right)
\end{equation}
and
\begin{equation} \label{pot2M}
\oint B_\varphi d\varphi = \frac{2 \pi}{\widehat{E}_M}  \left( 1 -
\frac{2}{\widehat{E}_M^2 \rho_M^2} \right).
\end{equation}
The dilaton at the boundary is
\begin{eqnarray} \label{dilatE}
e^{-2\phi} &=& e^{-2\phi_0} \frac{\Lambda(\xi_3)
\widehat{B}_E^2}{\Psi(\xi_3) \widehat{E}_E^2} \left[ 1 + \frac{2
\Sigma A}{1-\Sigma^2 A^2 \xi_3^2} \frac{4 F(\xi_3) L^2}{G'(\xi_3) A^2
\rho_E^2} \right. \\ \nonumber && \left. + \frac{2}{\widehat{B}_E^2
\rho_E^2} - \frac{2}{\widehat{E}_E^2 \rho_E^2} \right]
\end{eqnarray}
in the $U(1)^2$ Ernst solution, and
\begin{equation} \label{dilatM}
e^{-2\phi} = \frac{\widehat{B}_M^2}{\widehat{E}_M^2} \left( 1 +
\frac{2}{\widehat{B}_M^2 \rho_M^2} -
\frac{2}{\widehat{E}_M^2 \rho_M^2} \right)
\end{equation}
in the $U(1)^2$ Melvin solution. Now $e^{\phi_0}, \widehat{B}_M$ and
$\widehat{E}_M$ are given by (\ref{expang}) and (\ref{subl3}), and we may
see that we can match the proper length of the boundary, the integrals
of the gauge fields and the dilaton if we also take
\begin{equation} \label{rho}
\rho_M = \rho_E \left\{ 1+ \frac{1}{\rho_E^2} \frac{F(\xi_3)
L^2}{G'(\xi_3) A^2} \left[ \frac{H''(\xi_3)}{H'(\xi_3)} - \frac{2
F'(\xi_3)}{F(\xi_3)} \right] \right\}.
\end{equation}

This implies that the difference in horizon area is
\begin{eqnarray} \label{diffha}
\Delta {\cal A} &=& -\frac{4 \pi L^2 F(\xi_3)}{G'(\xi_3) A^2} \left[
\frac{1}{\xi_4 - \xi_3} + \frac{H''(\xi_3)}{2H'(\xi_3)} -
\frac{F'(\xi_3)}{F(\xi_3)} \right]
\\ \nonumber &=& - \frac{4 \pi L^2 F(\xi_3)}{G'(\xi_3) A^2} \left[
\frac{(\xi_2-\xi_1)}{(\xi_3-\xi_2)(\xi_3-\xi_1)} + \frac{2}{(\xi_3 -\xi_1)}
-\frac{F'(\xi_3)}{F(\xi_3)} \right].
\end{eqnarray}
For the extreme case, we therefore have
\begin{equation}
 - \frac{1}{4} \Delta {\cal A} =  \frac{2 \pi L^2 F(\xi_3)}{ G'(\xi_3)
A^2} \left[ \frac{1}{\xi_3- \xi_1} - \frac{F'(\xi_3)}{2 F(\xi_3)} \right],
\end{equation}
while for the non-extreme case, we have
\begin{eqnarray}
-\frac{1}{4} (\Delta {\cal A} + {\cal A}_{bh}) &=& \frac{\pi
L^2 F(\xi_3)}{G'(\xi_3) A^2} \left[  \frac{2}{\xi_3- \xi_1} -
\frac{F'(\xi_3)}{
F(\xi_3)} \right. \\ \nonumber && \left. +
\frac{(\xi_2-\xi_1)}{(\xi_3-\xi_2)(\xi_3-\xi_1)} - \frac{
F(\xi_2) (\xi_4-\xi_3)}{F(\xi_3) (\xi_3-\xi_2) (\xi_4 - \xi_2)}
\right] \\ \nonumber &=&  \frac{2 \pi L^2 F(\xi_3)}{ G'(\xi_3)
A^2} \left[ \frac{1}{\xi_3- \xi_1} - \frac{F'(\xi_3)}{2 F(\xi_3)} \right],
\end{eqnarray}
where I have used the instanton condition (\ref{instcond}) to cancel the
last two terms. Thus, I deduce that the action must be
\begin{equation} \label{actionres}
I_{b} =  \frac{2 \pi L^2 F(\xi_3)}{ G'(\xi_3)
A^2} \left[ \frac{1}{\xi_3- \xi_1} - \frac{F'(\xi_3)}{2 F(\xi_3)} \right]
\end{equation}
in both cases.  This answer agrees with the action of the Ernst
solution found in \cite{ggs,entarea} when $\Sigma =0$ (which implies
$F(\xi)=1$), as it should. It also reduces to the answer for the
action of the dilaton Ernst solution found in \cite{dggh,entarea} when
either $Q=0$ or $P=0$. Thus, this result is consistent with the
previously-obtained results.

The point-particle limit is $r_+ A \ll 1$, as the black hole
becomes small on the scale set by the acceleration in this limit. In
this limit, both the extreme and non-extreme black holes satisfy $r_+
\approx r_-$ \cite{2u1}. When $r_+ A \ll 1$, the action reduces to
\begin{equation}
  I_b \approx \frac{\pi r_-}{A} \approx \frac{\pi M^2}{BP + EQ},
  \label{lowact}
\end{equation}
where I have used Newton's law in the second step. The pair creation
rate is $e^{-I_b}$, so we recover the Schwinger result (generalised
to the case of two gauge fields) in this limit, as we would
expect. That is, we find that small black holes are pair created at
the same rate (to leading order) as we would expect for some
hypothetical particles carrying the same mass and charges. In
particular, the pair creation rate will be very small for realistic
fields, as we must have $M> M_{pl}$ for this semi-classical
approximation to be valid. Because of the number of
parameters involved, it is difficult to say anything more about the
general behaviour of this action, but the qualitative agreement with
\cite{entarea} is remarkable, given the much more complicated nature
of this solution, and the presence of twice as many free parameters.

\section{Acknowledgements}

I am happy to acknowledge many helpful conversations with friends and
colleagues, particularly Stephen Hawking, who I also thank for his
comments on an earlier draft of this paper. I thank the Association of
Commonwealth Universities and the Natural Sciences and Engineering
Research Council of Canada for financial support.

\end{document}